\documentstyle[12pt,amssymb]{article}
\def\sg{\sqrt{-g}}
\def\pa{\partial}
\def\det{\mathop{\rm det}\nolimits}
\def\gab{g^{\alpha\beta}}
\def\dom{\delta_\omega}
\def\pap{2\pi\alpha'}

\title{The Green--Schwarz Superstring in Extended Configuration Space
and Infinitely Reducible First Class Constraints Problem}
\author{A.A. Deriglazov and A.V. Galajinsky\thanks{E-mail:
deriglaz@phys.tsu.tomsk.su;galajin@phys.tsu.tomsk.su}}
\date{Department of Theoretical Physics, Tomsk State University, 634050
Tomsk, Russia}
\begin{document}
\maketitle
\begin{abstract}
The Green--Schwarz superstring action is modified to include some set of
additional (on-shell trivial) variables. A complete constraints system
of the theory turns out to be reducible both in the original and in
additional variable sectors. The initial $8s$ first class constraints
and $8c$ second class ones are shown to be unified with $8c$ first and
$8s$ second class constraints from the additional variables sector,
resulting with $SO(1,9)$-covariant and linearly independent constraint sets.
Residual reducibility proves to fall on second class constraints only.
\end{abstract}


\section{Introduction}
The general recipe of covariant quantization of dynamical systems subject to
reducible first and second class constraints was developed in Refs.
1--3. ``Ghosts for ghosts'' mechanism [1, 2] was proposed to balance
correct dynamics on the one hand and manifest covariance on the
another. Application of the scheme turned out to be remarkably
successful for certain cases. The antisymmetric tensor field [1],
chiral superparticle [4], high superspin theories [5] seem to be the
most interesting examples.

However, in the general case there may arise an infinite tower of extra
ghost variables, what makes the expression for effective action formal.
The superparticle [6] and superstring [7] models appeared to be the
first (and, actually, the most important) examples of such a kind. A
complete constraints system of the theories in the Hamiltonian
formalism includes fermionic constraints\footnote{We discuss mostly
$N=1, D=10$ case for that covariant quantization is the principal
problem.} which, being a mixture of 8 first class and 8 second class
ones ({\bf 8}s and {\bf 8}c-representations of $SO(8)$-little group
respectively), lie in the minimal spinor representation of the Lorentz
group. The latter fact means that covariant irreducible separation of
the constraints is impossible in the original phase space [8]. However,
one can realize reducible split by making use of covariant projectors
known for the superparticles [9--11] and superstring [12]. Introduction of 16
covariant primary ghosts to the (reducible) first class constraints
implies 16 secondary ones etc. There arises an infinite tower of extra
ghost variables. The Lagrangian analog of the situation is infinitely
reducible Siegel symmetry [13], with spinor parameters from which only
half is essential on-shell. Note that within the framework of the
alternative twistor-harmonic approach [14], the fermionic constraints
can be separated in covariant and irreducible manner due to the
``bridge nature'' of the harmonic variables. This formalism, however,
is essentially Hamiltonian and the reparametrization invariance of the
original Green--Schwarz theory turns out to be broken in the modified
version [14].

Reformulation of the BFV-procedure which do not involve explicit
separation of constraints was presented in Refs. 15--17. However, as
was shown in Ref. 11, application of the scheme for concrete models may
conflict with manifest Poincar\'e covariance.

In this paper we propose an alternative approach to the infinitely
reducible constraints problem of $D=10, N=1$
Green--Schwarz superstring (GSS). The basic idea is to introduce
additional pure gauge fermionic degrees of freedom subject to {\it
reducible} constraints like that of the GSS. We choose these
constraints to be a pair of Majorana--Weyl spinors with the following
structure:\footnote{The total number of constraints is sufficient to
suppress just one canonical pair of variables.}

(i) The first of them is a mixture of 8 first class and 8 second
class constraints, which are required to lie in {\bf 8}c and {\bf 8}s
irreducible representations of $SO(8)$ group, respectively.

(ii) The second spinor contains only 8 linearly-independent components
being second class constraints.

Splitting further all the fermionic constraints of the problem in
covariant and reducible manner (by making use of covariant projectors
[11, 12]) one can combine the original fermionic first class
constraints of the GSS with the first class ones from the additional
variables sector into one irreducible set (what corresponds to $\bar
8s\oplus 8c$-representation of $SO(8)$ or Majorana-Weyl spinor of
$SO(1,9)$. Analogously, the second class constraints from the
additional variables sector can be unified with the original second
class ones resulting with covariant and irreducible constraint. For the
model concerned, the resulting constraint system turns out to be
completely equivalent to the initial one. Thus, the reducible fermionic
first class constraints of the GSS become irreducible in the modified
theory. The infinite tower of extra ghost variables, that corresponds
to the first class constraints in the original formulation of the
superstring, will no appear in the new version. The Lagrangian which
reproduce the scheme described above is our main result.

The paper is organized as follows. In Sec. 2 the Green-Schwarz action
is modified to include some set of additional variables. The local
symmetries of the model are investigated. A complete canonical analysis
of the theory is carried out in subsec. 3.1. Classical equivalence
of the modified and original superstrings is established in subsec.
3.2. We do this by imposing gauge conditions for all first class
constraints in the problem. Dynamics in the physical variables
sector proves to coincide with that of the GSS. Note that all the
gauge conditions can be imposed in covariant manner, excepting the standard
light-cone gauge conditions corresponding to the super-Virasoro constraints.
In Sec. 4 explicitly covariant separation of the constraints is
realized. The infinitely reducible first class constraints problem is
resolved. Concluding remarks are presented in Sec. 5. Appendix A
contains our conventions and a brief description of $SO(8)$-formalism
used in the work. Appendix B includes essential Poisson brackets of the
constraints involved.

\section{Action and local symmetries}

The action functional to be examined is of the form
\begin{equation}
S=S_{\rm GS}+S_{\rm add},
\end{equation}
where
\begin{eqnarray*}
S_{\rm GS}&=&\int d\tau d\sigma\Big\{-\frac 1{4\pi\alpha'}\sg
g^{\alpha\beta}{\Pi^m}_\alpha\Pi_{m\beta}-\frac 1{2\pi\alpha'}
\epsilon^{\alpha\beta}\pa_\alpha X^mi\Theta\Gamma_m\pa_\beta\Theta\Big\},\\
S_{\rm add}&=&\int d\tau d\sigma\Big\{-\frac 12 \epsilon^{\alpha\beta}
\Lambda_m\big(\pa_\alpha {A^m}_\beta-\pa_\beta{A^m}_\alpha-\cr
&-&\pa_\alpha\Theta\Gamma^m\chi_\beta+i\pa_\beta\Theta\Gamma^m
\chi_\alpha +i\chi_\alpha\Gamma^m\chi_\beta\big)-\Phi\Lambda^2\Big\},
\end{eqnarray*}
and ${\Pi^m}_\alpha\equiv\pa_\alpha X^m-i\Theta\Gamma^m\pa_\alpha\Theta$,
$\sg\equiv\sqrt{-\det g_{\alpha\beta}}$. The first term in Eq. (1) is
the Green-Schwarz action [7], the second term is the action of
additional variables. All the variables are treated on equal footing.
The Latin indices are designed for target manifold tensors, the Greek
ones are set for worldsheet tensors (for instance, ${\chi^A}_\alpha$ is
$D=10$ Lorentz spinor and $D=2$ worldsheet vector). Statistics of the
fields corresponds to their tensor structure, i.e., $X^m$,
$g^{\alpha\beta}$, ${A^m}_\alpha$, $\Lambda^m$, $\Phi$ are bosons,
while $\Theta^A$, ${\chi^A}_\alpha$ are fermions. The matrix
$\epsilon^{\alpha\beta}$ is chosen in the form $\epsilon^{\alpha\beta}
=-\epsilon^{\beta\alpha}$, $\epsilon^{01}=-1$.

Since the $S_{\rm add}$ contains only derivatives of the $\Theta$, the
modified superstring is invariant under standard global supersymmetry
transformations.

Local symmetries of the theory, except the standard reparametrizations
of worldsheet and the Weyl transformations, include a modification of
the Siegel transformations\footnote{To check $k$-invariance of the action
it is necessary to use to the Fierz identity $\Gamma^m_{A(B}\Gamma^m_{CD)}=0$
and the property of the $P^\pm$ projectors: $P^{\pm\alpha\gamma}
P^{\pm\beta\sigma}=P^{\pm\beta\gamma}P^{\pm\alpha\sigma}$. Note as well
that the $k$-symmetry is reducible. The following transformation of
parameters $k'_\beta=k_\beta+\Pi_{n\gamma}\Gamma^nP^{-\gamma\sigma}
\varkappa_{\sigma\beta}$, with the $\varkappa_{\sigma\beta}$ being an arbitrary
function, does not change Eq. (2) (modulo equations of motion), what
means linear dependence of generators of the transformations.}
\begin{eqnarray}
&& \delta_k\Theta=2i\Pi_{m\alpha}\tilde\Gamma^mk^{-\alpha},\cr
&& \delta_kX^m=i\Theta\Gamma^m\delta_k\Theta,\cr
&& \delta_k(\sg \gab)=16\sg P^{-\alpha\gamma}(\pa_\gamma\Theta
k^{-\beta}),\\
&& \delta_k\chi_\alpha=\pa_\alpha(\delta_k\Theta),\cr
&& \delta_k{A^m}_\alpha=i\Theta\Gamma^m\pa_\alpha(\delta_k\Theta),
\nonumber
\end{eqnarray}
where
\[ P^{\pm\alpha\beta}\equiv \frac 12\Big(\gab\pm
\frac{\epsilon^{\alpha\beta}}{\sg}\Big), \qquad k^-\equiv P^-k, \]
and a set of new symmetries acting on the additional variables
subspace. Here we list them with brief comments.

There is a pair of bosonic symmetries with $D=10$ vector $\xi^m$ and
$D=2$ vector $\mu_\alpha$ parameters
\begin{eqnarray}
&& \delta_\xi {A^m}_\alpha=\pa_\alpha\xi^m,\\
&& \begin{array}{l} \delta_\mu{A^m}_\alpha=\Lambda^m\mu_\alpha,\\
\delta_\mu\Phi=-\displaystyle\frac 12 \epsilon^{\alpha\beta}\pa_\alpha
\mu_\beta,\end{array}
\end{eqnarray}
which mean that the fields ${A^m}_\alpha$ and $\Phi$ may be gauged away.
Note that the system (3), (4) is reducible. This can easily be seen by
taking $\xi^m=\Lambda^m\nu$, $\mu_\beta=\pa_\beta\nu$, where $\nu$ is
an arbitrary function. With such a choice $(\delta_\xi-\delta_\mu)|_{\rm
on-shell}\equiv0$, what means functional dependence of generators of
the transformations. In addition to the transformations (3) and (4) the
action (1) possesses the following fermionic symmetries
\begin{eqnarray}
&& \begin{array}{l} \delta_{s^+}\chi_\alpha=\Lambda_n\tilde\Gamma^n
{s^+}_\alpha,\\
\delta_{s^+}\Phi=\epsilon^{\alpha\beta}i(\pa_\alpha\Theta-\chi_\alpha)
{s^+}_\beta,\end{array}\\
&& \begin{array}{l} \delta_{s^-}\chi_\alpha=\Lambda_n\tilde\Gamma^n
{s^-}_\alpha,\\
\delta_{s^-}\Phi=i\epsilon^{\alpha\beta}(\pa_\alpha\Theta-\chi_\alpha)
{s^-}_\beta.\end{array}
\end{eqnarray}
The symmetries (5) and (6) are reducible. The transformation of
parameters, under which Eqs. (5) and (6) are invariant (modulo equations
of motion), is of the form
\[ s'_\alpha=s_\alpha+\Lambda_n\Gamma^n\varkappa_\alpha \]
with an arbitrary $\varkappa_\alpha$. It is interesting to note that
the reducible symmetries (2) and (6) can be replaced by one
irreducible symmetry if one supposes that $(\Lambda\Pi_\alpha)^2\ne0$.
Actually, consider the following transformation
\begin{equation}
\begin{array}{l} \dom\Theta=2i\Pi_{m\alpha}\tilde\Gamma^m
\omega^{-\alpha},\\
\dom X^m=i\Theta\Gamma^m\dom\Theta,\\
\dom(\sg\gab)=16\sg P^{-\alpha\gamma}(\pa_\gamma\Theta\omega^{-\beta}),\\
\dom\chi_\alpha=\pa_\alpha(\dom\Theta)+\Lambda_m\tilde\Gamma^m
{\omega^-}_\alpha,\\
\dom{A^m}_\alpha=i\Theta\Gamma^m\pa_\alpha(\dom\Theta),\\
\dom\Phi=i\epsilon^{\alpha\beta}(\pa_\alpha\Theta-\chi_\alpha)
{\omega^-}_\beta,\end{array}
\end{equation}
which is a formal sum of Eqs. (2) and (6) with $k^-=s^-\equiv\omega^-$.
Two remarks concerning this symmetry are relevant. First, it is
straightforward to check that there is no a transformation of
parameters which leaves Eq. (7) invariant, i.e., all 16 parameters are
effective on-shell. Secondly, the original $k^-$ and $s^-$
transformations (each of them has 8 essential parameters on-shell) can
be extracted from Eq. (7) by taking
\begin{equation}
\begin{array}{l} \omega_{1\beta}=\displaystyle\frac 1{(\Lambda\Pi)^2}
(\Lambda\Pi_\sigma)P^{+\sigma\gamma}\Lambda_n\Gamma^n\tilde\Gamma^m
\Pi_{m\gamma}k_\beta,\\
\omega_{2\beta}=\displaystyle\frac 1{(\Lambda\Pi)^2}(\Lambda\Pi_\sigma)
P^{+\sigma\gamma}\Pi_{n\gamma}\Gamma^n\tilde\Gamma^m \Lambda_m
s_\beta.\end{array}
\end{equation}

Equations of motion for the theory (1) are of the form
\begin{eqnarray*}
&& \begin{array}{l} {\Pi^m}_\alpha\Pi_{m\beta}=\displaystyle\frac 12
g_{\alpha\beta} g^{\gamma\delta}{\Pi^m}_\gamma\Pi_{m\delta},\\
\pa_\beta(\sg g^{\beta\alpha}\pa_\alpha X^m + 2\sg P^{-\beta\alpha}
i\Theta\Gamma^m\pa_\alpha\Theta)=0,\\
\Pi_{m\alpha}\Gamma^mP^{-\alpha\beta}\pa_\beta\Theta=0;\end{array}
\qquad\qquad\qquad\quad (9.a)\\[1ex]
&& \begin{array}{l} \Lambda^2=0,\\
\epsilon^{\alpha\beta}(\pa_\alpha{A^m}_\beta-\pa_\beta{A^m}_\alpha
-i\pa_\alpha\Theta\Gamma^m\chi_\beta+\\
\qquad +i\pa_\beta\Theta\Gamma^m\chi_\alpha+i\chi_\alpha\Gamma^m
\chi_\beta)+4\Phi\Lambda^m=0,\\
\pa_\alpha\Lambda_m=0,\\
i(\chi_\alpha-\pa_\alpha\Theta)\Gamma^m\Lambda_m=0.\end{array} \qquad
\qquad\qquad\qquad\quad(9.b)
\end{eqnarray*}
\stepcounter{equation}
Note that equations (9a) are just the Green--Schwarz superstring equations.
In the light-cone gauge this system reduces to $\Box X^i=0$,
$\pa_-\Theta^a=0$, where $i$ and $a$ are, respectively, vector and
spinor indices of $SO(8)$ group. It turns out that there are no more
dynamical degrees of freedom in the question. We will prove this fact
in the next section by passing to the Hamiltonian formalism and
imposing all gauge conditions.

\section{Canonical formalism}
\subsection{Dirac procedure}

Denoting momenta conjugate to the variables\footnote{Within the
framework of canonical formalism it is useful to make an invertible
change of variables [33]
\[ g^{00}, g^{01}, g^{11} \to N=-\frac 1{\sg g^{00}}, N_1=-\frac{g^{01}}
{g^{00}}, g^{11}, \]
where $\sg=\sqrt{-\det\,g_{\alpha\beta}}$. In terms of the new
variables finding of secondary constraints becomes evident.}
($X^m$, $\Theta^B$, $N$, $N_1$, $g^{11}$, $\Lambda^m$, ${A^m}_\alpha$,
${\chi^B}_\alpha$, $\Phi$) as ($\pi_m$, $P_{\theta B}$, $P_N$, $P_{N_1}$,
$P_g$, $\pi_{\Lambda m}$, $\pi_{Am}{}^\alpha$, $P_{\chi B}^\alpha$, $\pi_\Phi$) one gets
\begin{eqnarray}
&& \pi_m=\displaystyle\frac 1{\pap}\Big(\frac 1N \Pi_{m0}-\frac{N_1}{N}
\Pi_{m1}+i\Theta\Gamma_m\pa_1\Theta\Big),\cr
&& \tilde L\equiv P_\theta+i\Theta\Gamma^m\Big(\pi_m+\displaystyle \frac
1{\pap}\Pi_{m1}\Big)-i\chi_1\Gamma^m\Lambda_m\approx0;\cr
&& \pi_{\Lambda m}\approx0, \qquad \pi_{Am}{}^0\approx0, \qquad
\pi_{Am}{}^1-\Lambda_m\approx0,\\
&& {P_\chi}^0\approx0, \qquad {P_\chi}^1\approx0, \qquad
\pi_\Phi\approx0,\cr
&& P_N\approx0, \qquad P_{N_1}\approx0, \qquad
P_g\approx0,\nonumber
\end{eqnarray}
where $\pa_1\equiv\pa/\pa\sigma$. The first equation in Eq. (10)
determines $\pa_0X^m$ as a function of the other canonical variables.
The remaining equations are primary constraints.

The canonical Hamiltonian is given by
\begin{eqnarray}
\lefteqn{H=\int d\sigma\Big\{N_1(\hat\pi\Pi_1)+N\frac 12\Big(\pap
\hat\pi^2 + \frac 1{\pap}{\Pi_1}^2\Big)-}\cr
&& -A_0\pa_1\Lambda+i(\chi_1-\pa_1\Theta)\Gamma^m\Lambda_m\chi_0+
\Phi\Lambda^2+\tilde L\lambda_\theta+\cr
&& +P_N\lambda_N+P_{N_1}\lambda_{N_1}+P_g\lambda_g+
\pi_\Lambda\lambda_1+\pi_A{}^0\lambda_2+\cr
&& +({\pi_A}^1-\Lambda)\lambda_3+{P_\chi}^0\lambda_4+{P_\chi}^1
\lambda_5+\pi_\Phi\lambda_6\Big\},
\end{eqnarray}
where
\[ \hat\pi^n\equiv\pi^n-\frac 1{\pap} i\Theta\Gamma^n\pa_1\Theta \]
and $\lambda_\theta, \lambda_N, \lambda_{N_1}, \lambda_g,
\lambda_1-\lambda_6$ are Lagrange multipliers corresponding to the
primary constraints. The preservation in time of the primary
constraints implies the secondary ones
\begin{eqnarray}
&& \hat\pi_m{\Pi^m}_1\approx0, \qquad \frac 12\Big(\pap \hat\pi^2+
\frac 1{\pap}{\Pi_1}^2\Big)\approx0,\cr
&& \Lambda^2\approx0, \qquad (\chi_1-\pa_1\Theta)\Gamma^m\Lambda_n
\approx0,\\
&& \pa_1\Lambda^m\approx0,\nonumber
\end{eqnarray}
and conditions on the Lagrange multipliers
\begin{eqnarray*}
&& {\lambda_1}^m=0, \qquad\qquad\qquad\qquad\qquad\qquad\qquad\qquad\qquad
\qquad\qquad\quad~~ (13.a)\\
&& \lambda_{3\,m}=-i\chi_1\Gamma_m\lambda_\theta+\pa_1A_{m\,0}
+i(\chi_1-\pa_1\Theta)\Gamma_m\chi_0+2\Phi\pi_{A\,m}{}^1, \qquad~ (13.b)\\
&& \pi_{Am}{}^1\Gamma^m(\lambda_\theta-\chi_0)=0, \qquad\qquad\qquad\qquad\qquad
\qquad\qquad\qquad\qquad (13.c)\\
&& 2i\Gamma^m\Big(\hat\pi_m+\frac 1{\pap}\Pi_{m\,1}\Big)
\big(\lambda_\theta-(N+N_1)\pa_1\theta\big)- {}\\
&& \qquad -i\Gamma^m\pi_{Am}{}^1(\lambda_5-\pa_1\chi_0)=0. \qquad\qquad\qquad
\qquad\qquad\qquad\qquad (13.d)
\end{eqnarray*}
\stepcounter{equation}
Equations (13.c) and (13.d) are sufficient to determine
$\lambda_\theta$. Actually, multiplying Eq. (13.c) by $\tilde\Gamma^n
\Big(\hat\pi_n+\frac 1{\pap}\Pi_{n\,1}\Big)$ and Eq. (13.d) by
$\tilde\Gamma^n\pi_{An}{}^1$ and then taking the sum one gets
\begin{eqnarray}
\lefteqn{\lambda_\theta=\displaystyle\frac 1{2{\pi_A}^1\Big(\hat\pi
+\frac 1{\pap}\Pi_1\Big)}\Big\{\tilde\Gamma^n\Big(\hat\pi_n+ \frac
1{\pap}\Pi_{n\,1}\Big)\Gamma^m\pi_{Am}{}^1\chi_0+}\cr
&& +\tilde\Gamma^m\pi_{Am}{}^1\Gamma^n\Big(\hat\pi_n+\frac 1{\pap}
\Pi_{n1}\Big)(N+N_1)\pa_1\Theta\Big\}
\end{eqnarray}
provided that ${\pi_A}^1\Big(\hat\pi+\frac 1{\pap}\Pi_1\Big)\ne0$
on-shell. The latter condition can always be realized by choosing
appropriate gauge fixing conditions and initial data to the equations
of motion.\footnote{The constraints $\pa_1\pi_{An}{}^1\approx0$,
$({\pi_A}^1)^2\approx0$ together with the equation of motion
$\pa_0\pi_{An}{}^1=0$ imply $\pi_{Am}{}^1=n_m$, where $n_m$ is a
constant null-vector (initial data). Choosing the initial data to be
$n^m=(n^0,0,\dots,0,n^0)$, $n^0\ne0$, and imposing standard light-cone
gauge conditions (see subsec. 3.2) $X^-=\alpha'\tau p^-$, $\pi^-=\frac
1{2\pi} p^-$, where $p^-\ne0$ is a complete momentum of the
superstring, one gets
\[ {\pi_A}^1\Big(\hat\pi+\frac 1{\pap}\Pi_1\Big)=-\frac 1{2\pi}n^+
\Big(p^--\frac 2{\alpha'}i\theta\Gamma^-\pa_1\theta\Big)\ne0. \]}
Inserting further Eq. (14) into Eq. (13.d) one finds
\begin{equation}
\pi_{An}{}^1\Gamma^n(\lambda_5-\pa_1\chi_0)=0.
\end{equation}
Equation (15) determines half of the $\lambda_5$, that can easily be
seen by passing to the $SO(8)$-formalism. In $SO(8)$-notations the
condition (15) reads (see Appendix A)
$$
(\lambda_{5\dot a}-\pa_1\chi_{0\dot a})-\frac 1{\sqrt{2}{\pi_A}^{+1}}
\gamma^i_{\dot aa}\pi_{Ai}{}^1(\lambda_{5a}-\pa_1\chi_{0a})=0.
$$
It is straightforward to check further the primary constraints are
identically conserved in time if Eqs. (12) and (13) hold.
Thus, there are no more constraints in the problem.

To separate the original constraints of the theory into first and
second class, consider the equivalent constraints system
\begin{eqnarray*}
&& \pi_{\Lambda n}\approx0, \qquad \Lambda_n-\pi_{An}{}^1\approx0,
\qquad\qquad\qquad\qquad\qquad\qquad\qquad\quad~ (16.a)\\
&& \begin{array}{l} \pi_{An}{}^0\approx0, \qquad \pi_\Phi\approx0,\\
{P_\chi}^0\approx0, \qquad P_N\approx0,\\
P_{N_1}\approx0, \qquad P_g\approx0,\\
({\pi_A}^1)^2\approx0, \qquad \pa_1\pi_{An}{}^1\approx0,\end{array}
\qquad\qquad\qquad\qquad\qquad\qquad\qquad\quad~ (16.b)\\
&& \hat\pi_n{\Pi^n}_1+L\pa_1\Theta\approx0, \qquad \displaystyle\frac 12\Big(
\pap \hat\pi^2+\frac 1{\pap}{\Pi_1}^2\Big)+L\pa_1\Theta\approx0,
\qquad (16.c)\\
&& L\equiv P_\theta+i\Theta\Gamma^m\Big(\pi_m+\displaystyle\frac 1{\pap}
\Pi_{m1}\Big)-i\pa_1\Theta\Gamma^m\pi_{Am}{}^1-\pa_1{P_\chi}^1\approx0,
\quad (16.d)\\
&& {P_\chi}^1\approx0, \qquad (\chi_1-\pa_1\Theta)\Gamma^n\pi_{An}{}^1
\approx0. \qquad\qquad\qquad\qquad\qquad\qquad~ (16.e)
\end{eqnarray*}
\stepcounter{equation}
The constraints (16.a) are second class. The constraints system (16.b),
(16.c) is first class. Among 16 fermionic constraints (16.d) half is
first class and another half is second class (see Sec. 4). Analogously,
the first equation in Eq. (16.e) contains 8 first class and 8 second
class constraints while the latter implies 8 linearly independent
second class constraints (see Sec. 4). The essential Poisson brackets
of the constraints (16) are gathered in the Appendix B.

Note that among 1+10 constraints $({\pi_A}^1)^2\approx0$,
$\pa_1{\pi_{An}}^1\approx0$ only 10 are functionally independent as a
consequence of the identity $\pa_1({\pi_A}^1)^2-2{\pi_A}^{n1}
\pa_1{\pi_{An}}^1\equiv0$. Independent constraints can be extracted in
the light-cone basis as
follows:
\[ {\pi_A}^{-1}-\frac 1{2{\pi_A}^{+1}}{\pi_A}^{i1}{\pi_{Ai}}^1\approx0,
\qquad \pa_1{\pi_A}^{+1}\approx0, \qquad \pa_1{\pi_{Ai}}^1\approx0. \]
As was mentioned above, it is impossible to separate 8 first and 8
second class constraints, being combined in the $L$, in covariant and
irreducible manner. For the model concerned, covariant projectors
into first and second class constraints are constructed in Sec. 4.

An explicit counting the degrees of freedom shows that there are 16
bosonic and 8 fermionic phase space degrees of freedom in the model
that just coincides with the number of degrees of freedom in the
Green--Schwarz theory. Note as well that, after use of the Dirac
algorithm, there remained 1+2+10+1 bosonic and 16+8 fermionic undefined
Lagrange multipliers. Since the local symmetries, considered in Sec. 2,
have just this number of parameters being independent on-shell, we
conclude that they exhaust all the essential Lagrangian symmetries of
the model.

\subsection{Gauge fixing and physical dynamics}

In imposing gauge fixing conditions two criteria should be satisfied
[18]. First, the Poisson bracket of original first class
constraints and gauges must be an invertible matrix when restricted to
constraints and gauges surface. Secondly, gauge conditions are
to be consistent with equations of motion, i.e., there must no
appear new constraints from condition of preservation in time of the
gauges.\footnote{In the general case one can admit new constraints if
they will further be treated as gauge conditions for some of the original
first class constraints.} With this remark, consider first gauge conditions
fixing all the undefined Lagrange multipliers in the theory. The following
equations:
\begin{eqnarray*}
&& N\approx1, \quad N_1\approx0, \quad g^{11}\approx1, \quad
{A^m}_0\approx0, \quad \Phi\approx1/2, \qquad\qquad\quad~ (17.a)\\
&& \chi_0\approx0, \quad (\chi_1-\pa_1\Theta)\Gamma^n\Big(\hat\pi_n+
\displaystyle\frac 1{\pap}\Pi_{n1}\Big)\approx0 \qquad\qquad\qquad\qquad\quad (17.b)
\end{eqnarray*}
\stepcounter{equation}
prove to be suitable for this goal. Preservation in time of the
gauges (17) yields
\begin{equation}
\begin{array}{l} \lambda_N=0, \qquad \lambda_{N_1}=0, \qquad
\lambda_g=0,\\
{\lambda_2}^m=0, \qquad \lambda_6=0, \qquad \lambda_4=0,\end{array}
\end{equation}
and
\begin{eqnarray}
&& (\lambda_5-\pa_1\lambda_\theta)\Gamma^n\Big(\hat\pi_n+
\displaystyle\frac 1{\pap}\Pi_{n1}\Big)+(\chi_1-\pa_1\Theta)\Gamma^n
\bigg(\frac 2{\pi\alpha'}i\pa_1\Theta\Gamma_n\lambda_\theta+\cr
&&\qquad +\pa_1\Big(\hat\pi_n+\displaystyle\frac 1{\pap}\Pi_{n1}\Big)\bigg)=0.
\end{eqnarray}
Taking into account that the constraint $(\chi_1-\pa_1\Theta)\Gamma^n
{\pi_{An}}^1\approx0$ together with the gauge
$(\chi_1-\pa_1\Theta)\Gamma^n \Big(\hat\pi_n+\frac 1{\pap}
\Pi_{n1}\Big)\approx0$ imply
\begin{equation}
\chi_1-\pa_1\Theta\approx0,
\end{equation}
while Eqs. (13.c), (15), (16.b), and (19) mean
\begin{equation}
\lambda_5-\pa_1\lambda_\theta=0
\end{equation}
one concludes that all the Lagrange multipliers have been fixed. In the
gauge chosen, the canonical variables $(N, P_N)$, $(N_1, P_{N_1})$,
$(g^{11},P_g)$, $({A^m}_0, \pi_{Am}{}^0)$, $(\chi_0, {P_\chi}^0)$,
$(\chi_1, {P_\chi}^1)$, $(\Phi, \pi_\Phi)$ are unphysical
and may be dropped after introducing the corresponding Dirac bracket.

Consider now gauge conditions to the remaining first class constraints.
\begin{enumerate}
\def\labelenumi{(\roman{enumi})}
\item $(\Theta, P_\theta)$-{\it sector}. There are 8 first class
constraints being nontrivially combined with 8 second class ones in Eq.
(16.d). In this case one can adopt the condition
\begin{equation}
\Gamma^n\pi_{An}{}^1\Theta\approx0
\end{equation}
or in the $SO(8)$-formalism (we write out only linearly independent
part of Eq. (22))
\begin{equation}
\Theta_{\dot a}-\frac 1{\sqrt{2}{\pi_A}^{+1}}\gamma^i_{\dot aa}
\Theta_a\pi_{Ai}{}^1\approx0.
\end{equation}
Because $\Gamma^n\pi_{An}{}^1\lambda_\theta\approx0$ (see Eq. (14)) the
gauge (22) is consistent with the equations of motion. It is
straightforward to check as well that the Poisson bracket of the first
class constraints being contained in the $L$ with the gauge (23) is an
invertible matrix.

After gauge fixing, the only dynamical variables in the sector are
$\Theta_a$. Taking into account that in the gauge chosen Eq. (14) takes
the form
\begin{equation}
\lambda_\theta=\pa_1\Theta
\end{equation}
one concludes that physical dynamics of the superstring (1) in the
odd-variables sector is described by
\[ \pa_0\Theta_a=\pa_1\Theta_a \]
or
\begin{equation}
\pa_-\Theta_a=0, \qquad a=1,\dots, 8,
\end{equation}
just as in the Green-Schwarz model.
\item $({A^m}_1, \pi_{Am}{}^1)$-{\it sector}. The constraints to be
discussed are of the form
\begin{equation}
({\pi_A}^1)^2\approx0, \qquad \pa_1{\pi_A}^{m1}\approx0.\label{26}
\end{equation}
The equations of motion in the sector read
\begin{equation}
\pa_0{\tilde A^m}_1=0, \qquad \pa_0\tilde\pi_A{}^{m1}=0,
\end{equation}
where canonical transformation
\[ {A^m}_1\to {\tilde A^m}_1={A^m}_1-\tau{\pi_A}^{m1}, \qquad
{\pi_A}^{m1}\to{\tilde\pi_A}^{m1}={\pi_A}^{m1} \]
has been made. Imposing then the following gauge condition to Eq. (\ref{26})
\[ \tilde A^m{}_1\approx0 \]
one concludes that there are no physical degrees of freedom in the sector.
\item $(X^m,\pi_m)$-{\it sector}.
There are super-Virasoro constraints (16.c) and equations of motion
\begin{equation}
\pa_0X^m=\pap \pi^m, \qquad \pa_0\pi^m=\frac 1{\pap}\pa_1\pa_1X^m.\label{30}
\end{equation}
In this case one can impose the standard light-cone gauges [19]
\begin{equation}
X^-=\alpha'\tau p^-, \qquad \pi^-=\frac 1{2\pi}p^-\label{31}
\end{equation}
where $p^-$ is the complete momentum of the superstring. It is easy to
check that the conditions (\ref{31}) are consistent with Eq. (\ref{30}). Making use
of Eq. (16.c) to express the variables $X^+$ and $\pi^+$ as
functions of other variables and taking into account the gauges (\ref{31})
one can conclude that the physical dynamics in the sector is described by
\begin{equation}
\pa_0X^i=\pap\pi^i, \qquad \pa_0\pi^i=\frac 1{\pap}\pa_1\pa_1X^i\label{32}
\end{equation}
or eliminating the $\pi^i$
\begin{equation}
\Box X^i=0, \qquad i=1,\dots,8.\label{33}
\end{equation}
Thus, physical dynamics of the superstring (1) is determined by Eqs.
(25) and (\ref{33}), what just coincides with the Green-Schwarz superstring
dynamics.
\end{enumerate}

\section{Covariant separation of constraints. Resolving the infinitely
reducible first class constraints problem}

In the previous sections we have modified the GSS so as to include a
set of additional pure gauge variables. In the extended phase space
covariant separation of constraints present no a special problem.
Actually, consider the following constraints system:
\begin{eqnarray*}
&& \pi_{An}{}^1\tilde\Gamma^n{P_\chi}^1\approx0,  \qquad\qquad\qquad\qquad
\qquad\qquad\qquad\qquad\qquad\qquad\quad (32.a)\\
&& b_n\tilde\Gamma^n{P_\chi}^1\approx0, \quad (\chi_1-\pa_1\Theta)
\Gamma^n\pi_{An}{}^1\approx0, \qquad\qquad\qquad\qquad\qquad\quad (32.b)\\
&& b_n\tilde\Gamma^nL\approx0, \qquad\qquad\qquad\qquad\qquad\qquad\qquad
\qquad\qquad\qquad\qquad\qquad (32.c)\\
&& \pi_{An}{}^1\tilde\Gamma^nL\approx0, \qquad\qquad\qquad\qquad\qquad
\qquad\qquad\qquad\qquad\qquad\qquad (32.d)
\end{eqnarray*}
\stepcounter{equation}
where $b^n\equiv\hat\pi^n+\frac 1{\pap}{\Pi^n}_1$, that is completely
equivalent to Eqs. (16.e) and (16.d) due to the condition $(b{\pi_A}^1)
\ne0$ (see subsec. 3.1). Passing to the $SO(8)$-formalism it is
straightforward to check now that Eq. (32.a) includes 8 linearly
independent first class constraints; Eq. (32.b) contains 8+8 independent
second class ones; Eqs. (32.c) and (32.d) imply 8 first class and 8
second class constraints, respectively. For instance, rewriting Eqs.
(32.a) and (32.b) into $SO(8)$ formalism one gets (see Appendix A)
\begin{eqnarray*}
&& \nu_a\equiv\sqrt{2}{\pi_A}^{+1}P_{\chi a}{}^1+\pi_{Ai}{}^1
{\gamma^i}_{a\dot a}P_{\chi\dot a}{}^1\approx0, \qquad\qquad\qquad\qquad\qquad\qquad (33.a)\\
&& \varphi_{\dot a}\equiv\sqrt{2}b^-P_{\chi\dot a}{}^1+b_i
{\gamma^i}_{\dot aa}P_{\chi a}{}^1\approx0, \qquad\qquad\qquad\qquad\qquad\qquad\qquad (33.b)\\
&& \psi_{\dot a}\equiv\sqrt{2}{\pi_A}^{+1}(\chi_{1\dot a}-
\pa_1\Theta_{\dot a})-(\chi_{1a}-\pa_1\Theta_a){\gamma^i}_{a\dot a}
\pi_{Ai}{}^1\approx0. \qquad\qquad\quad (33.c)
\end{eqnarray*}
\stepcounter{equation}
Evaluating then the Poisson brackets of the constraints
\begin{equation}
\{\varphi_{\dot a},\psi_{\dot b}\}\approx 2b^-{\pi_A}^{+1}\Big(
\delta_{\dot a\dot b}-\frac 1{2b^-{\pi_A}^{+1}}b_i\gamma^i_{\dot ac}
\gamma^j_{c\dot b}\pi_{Aj}{}^1\Big)\equiv\Delta_{\dot a\dot b},\label{36}
\end{equation}
(all other brackets vanish\footnote{Note that the constraints
$\nu_a\approx0$ and $\varphi_{\dot a}\approx0$ are equivalent to
$P_{\chi a}{}^1\approx0$, $P_{\chi\dot a}{}^1\approx0$ due to Eq.
(\ref{37}).}), and taking into account that the matrix in the right hand side
of Eq. (\ref{36}) is invertible
\begin{equation}
\Delta\tilde\Delta =1, \qquad \tilde\Delta_{\dot a\dot b}=- \frac
1{2(b{\pi_A}^1)}\Big(\delta_{\dot a\dot b}-\frac 1{2b^-{\pi_A}^{+1}}
\pi_{Ai}{}^1\gamma^i_{\dot ac}\gamma^j_{c\dot b}b_j\Big)\label{37}
\end{equation}
one concludes that the constraint $\nu_a$ is first class, while
$\varphi_{\dot a}$ and $\psi_{\dot b}$ are second class. Analogous
calculations can be performed for the constraints (32.c) and (32.d).
Thus, the constraint system (16.a)--(16.e) can be covariantly splitted
into first and second class.

Let us now discuss Eqs. (32). The remarkable observation is that the
reducible first class constraints (32.a), (32.c) (reducible second
class constraints (32.d) and the first equation in Eq. (32.b)) can be
combined to form irreducible constraints set. Actually, consider the
constraints system
\begin{eqnarray*}
&& \pi_{An}{}^1\tilde\Gamma^n{P_\chi}^1+b_n\tilde\Gamma^nL\approx0,
\qquad\qquad\qquad\qquad\qquad\qquad\qquad\qquad\quad (36.a)\\
&& b_n\tilde\Gamma^n{P_\chi}^1+\pi_{An}{}^1\tilde\Gamma^nL\approx0,
\qquad\qquad\qquad\qquad\qquad\qquad\qquad\qquad\quad (36.b)\\
&& (\chi_1-\pa_1\Theta)\Gamma^n\pi_{An}{}^1\approx0, \qquad\qquad\qquad
\qquad\qquad\qquad\qquad\qquad\qquad (36.c)
\end{eqnarray*}
\stepcounter{equation}
which is completely equivalent to the original one (Eqs. (16.d) and
(16.e)) due to the constraints $b^2\approx0$, $({\pi_A}^1)^2\approx0$.
The constraints (36.a) are first class and linearly independent.
Analogously, the constraints (36.b) are second class and irreducible.
The remaining constraints (36.c) are second class and reducible.

Thus, in the modified version (1) of the superstring, the fermionic
first class constraints form irreducible set. They will require only 16
covariant ghost variables (and 16 conjugate momenta) in constructing
the BRST-charge. The infinite tower of extra ghost variables, that
appeared for the GSS, will no arise in the modified model (the
remaining reducible first class constraints $({\pi_A}^1)^2\approx0$,
$\pa_1\pi_{An}{}^1\approx0$ are first stage of reducibility and can be
taken into account along the standard lines [1--3]).

Note that the operators extracting the first and second class constraints
from the initial mixed constraints system are not strict projectors. For
instance, $\Gamma^nb_n\tilde\Gamma^mb_m=b^2\approx0$. In certain cases [11],
however, it is more convenient to deal with the first and second class
constraints which were extracted by means of strict projectors. For the
model concerned the suitable projectors are of the form
\begin{equation}
P^\pm=\frac 12(1\pm K), \qquad K=\frac 1{2\sqrt{(b{\pi_A}^1)^2 -b^2
({\pi_A}^1)^2}}\tilde\Gamma^{[n}\Gamma^{m]} b_n\pi_{Am}{}^1.\label{39}
\end{equation}
In terms of the operators covariant (redundant) split of the
constraints looks as follows:
\begin{eqnarray*}
L^-\approx0, \qquad {P_\chi}^{+1}\approx0 && {\rm -~~first~class},\\
L^+\approx0, \qquad {P_\chi}^{-1}\approx0 && {\rm -~~second~class},
\end{eqnarray*}
where $L^\pm\equiv LP^\pm$, ${P_\chi}^{\pm1}\equiv{P_\chi}^1P^\pm$.
Generalization of Eqs. (36.a)--(36.c) reads
\begin{eqnarray}
&& L^-+{P_\chi}^{+1}\approx0,\cr
&& L^++{P_\chi}^{-1}\approx0,\\
&& (\chi_1-\pa_1\Theta)\Gamma^n\pi_{An}{}^1\approx0.\nonumber
\end{eqnarray}

\section{Final remarks}

In this work the infinitely reducible first class constraints problem
of the original GSS has been resolved. However, there still remains
(infinitely) reducible second class constraints in the question. As is
known, within the framework of the standard BFV-formalism first and second
class constraints are treated in different manner. First class
constraints contribute to the BRST-charge while second class ones
appear in the path integral measure [3, 20]. In this sense, the problem of
covariant quantization of the GSS reduces to constructing a correct
integral measure for the theory (1). The weak Dirac bracket
construction [11] appears to be suitable for this goal and this work is
in progress now.

Note as well that the proposed techniques can be directly applied to
modification of the superparticle (and superstring) due to Siegel [21,
22]. In that case, there are only 8 linearly independent fermionic
first class constraints in the initial formulation, and use of the
scheme will lead to the system with all fermionic constraints being
irreducible. After this, covariant quantization is straightforward and
the results will be present elsewhere.

\section*{Appendix A}

In this paper we use generalized notations in which two
inequivalent minimal spinor representations of the Lorentz group
(right-handed and left-handed Majorana--Weyl spinors) are distinguished
by position of its indices. We set lower index for the right-handed
spinor $\psi_A, \, (A=1,\dots, 16)$ and upper index for the left-handed
one $\psi^A$. The generalized $16\times16$ Dirac matrices are real,
symmetric, obeying the standard algebra
$$
\Gamma^m\tilde\Gamma^n+\Gamma^n\tilde\Gamma^m=2\eta^{mn}, \qquad
\eta^{nm}={\rm diag}\,(-,+,\dots).
\eqno{(A.1)}$$
In analyzing the constraint systems of the superparticle, superstring
models it is useful to represent a Majorana--Weyl spinor of
$SO(1,9)$-group as a Majorana one of $SO(8)$-group
$$
\Psi_A=(\psi_a,\bar\psi_{\dot a}), \qquad a,\dot a=1,\dots,8,
\eqno{(A.2)}$$
where indices $a,\dot a$ label two inequivalent minimal spinor
representations of $SO(8)$ group (8c- and 8s-representations,
respectively). This correspondence becomes evident in the
following basis of the $\Gamma$-matrices
$$
\begin{array}{cc}
\Gamma^0=\left(\begin{array}{cc} {\bf 1}_8 & 0\\ 0 & {\bf 1}_8
\end{array}\right), & \tilde\Gamma^0=\left(\begin{array}{cc} -{\bf 1}_8 & 0\\
0 & -{\bf 1}_8\end{array}\right),\\
\Gamma^i=\left(\begin{array}{cc} 0 & \gamma^i_{a\dot a}\\
\gamma^i_{\dot aa} & 0\end{array}\right), &\tilde\Gamma^i=\left(
\begin{array}{cc} 0 & \gamma^i_{a\dot a}\\
\gamma^i_{\dot aa} & 0\end{array}\right), \\
\Gamma^9=\left(\begin{array}{cc} {\bf 1}_8 & 0\\ 0 & -{\bf
1}_8\end{array}\right),
& \tilde\Gamma^9=\left(\begin{array}{cc} {\bf 1}_8 & 0\\
0 & -{\bf 1}_8\end{array}\right),\end{array}
\eqno{(A.3)}$$
where the $\gamma^i_{a\dot a}, \gamma^i_{\dot aa}\equiv(\gamma^i_{a\dot
a})^T$ are $SO(8)$ $\gamma$-matrices [19]
$$
\gamma^i_{a\dot a}\gamma^j_{\dot ab}+\gamma^j_{a\dot a}\gamma^i_{\dot ab}
=2\delta^{ij}\delta_{ab}, \quad i=1,\dots,8.
\eqno{(A.4)}$$
Let now $b^n$ be a light-like vector
$$
-2b^+b^-+b^ib_i=b^2=0.
\eqno{(A.5)}$$
The useful observation is that under the assumption (A.5) the equation
$$
b_n(\tilde\Gamma^n\psi)^A=0
\eqno{(A.6)}$$
determines only 8-linearly independent conditions.

Actually, rewriting Eq. (A.6) in the $SO(8)$ formalism one gets
\begin{eqnarray*}
&& \sqrt{2}b^+\psi_a+b_i\gamma^i_{a\dot a}\bar\psi_{\dot a}=0,
\qquad\qquad\qquad\qquad\qquad\qquad\qquad\qquad\qquad (A.7a)\\
&& \sqrt{2}b^-\bar\psi_{\dot a}+b_i\gamma^i_{\dot aa}\psi_a=0,
\qquad\qquad\qquad\qquad\qquad\qquad\qquad\qquad\qquad (A.7b)
\end{eqnarray*}
By virtue of Eqs. (A.4) and (A.5), the Eq. (A.7b) is a consequence of
Eq. (A.7a), provided that the standard light-cone assumption $b^+\ne0$
has been made.

\section*{Appendix B}
In this Appendix we list the essential Poisson brackets of the
constraints (16.a)--(16.e):
$$
\begin{array}{l}
\{L_A,L_B\}=2i\Gamma^n_{AB}\Big(\hat\pi_n+\displaystyle\frac 1{\pap}
\Pi_{n1}\Big)\delta(\sigma-\sigma');\cr
\{L_A,\hat\pi\Pi_1\}=-2i(\Gamma^n\pa_1\Theta)_A\Big(\hat\pi_n+
\displaystyle\frac 1{\pap}\Pi_{n1}\Big)\delta(\sigma-\sigma');\cr
\Big\{L_A,\displaystyle\frac 12\Big(\pap \hat\pi^2+\frac 1{\pap}
{\Pi_1}^2\Big)\Big\}=\cr
\qquad =-2i(\Gamma^n\pa_1\Theta)_A\Big(\hat\pi_n+
\displaystyle\frac 1{\pap}\Pi_{n1}\Big)\delta(\sigma-\sigma');\cr
\{\hat\pi\Pi_1,\hat\pi\Pi_1\}=2\hat\pi\Pi_1(\sigma)\pa_\sigma\delta+
\pa_1(\hat\pi\Pi_1)\delta;\cr
\Big\{\hat\pi\Pi_1, \displaystyle\frac 12\Big(\pap\hat\pi^2+
\frac 1{\pap}{\Pi_1}^2\Big)\Big\}=\Big(\pap\hat\pi^2+\frac 1{\pap}
{\Pi_1}^2(\sigma)\pa_\sigma\delta+\cr
\qquad + \pa_1\displaystyle\Big(\frac 12\Big(\pap\hat\pi^2+
\frac 1{\pap}{\Pi_1}^2\Big)\Big)\delta;\cr
\Big\{\displaystyle\frac 12\Big(\pap\hat\pi^2+\frac 1{\pap}{\Pi_1}^2
\Big),\frac 12\Big(\pap\hat\pi^2+\frac 1{\pap}{\Pi_1}^2\Big)\Big\}=\cr
\qquad =2(\hat\pi\Pi_1)(\sigma)\pa_\sigma\delta+\pa_1(\hat\pi\Pi_1)
\delta.\end{array}
\eqno{(B.1)}$$
In obtaining Eq. (B.1) the Fierz identity
$$
(\Gamma^n\psi)_A(\Gamma_n\varphi)_B+(\Gamma^n\psi)_B(\Gamma_n\varphi)_A
=\Gamma^n_{AB}(\varphi\Gamma_n\psi)
\eqno{(B.2)}$$
and the standard properties of the $\delta$-function
$$
\begin{array}{c} \pa_\sigma\delta=-\pa_{\sigma'}\delta,\\
F(\sigma')\pa_\sigma\delta=\pa_\sigma F(\sigma)\delta+F(\sigma)
\pa_\sigma\delta\end{array}
\eqno{(B.3)}$$
have been used.

\end{document}